\documentclass[aps,prd,twocolumn,showpacs,floatfix,superscriptaddress,groupedaddress,nofootinbib]{revtex4}  
\usepackage{graphicx}  
\usepackage{dcolumn}   
\usepackage{bm}        
\usepackage{amssymb, amsmath}   
\usepackage{hyperref}
\usepackage{tikz}
\usepackage{subfig}
\usepackage{url}

\newcommand{\bt}[1]{\mathbf{#1}}
\newcommand{\lya}{Lyman-$\alpha$ }
\newcommand{\lyam}{{\rm Ly}\alpha}
\newcommand{\nulya}{\nu_{\rm Ly\alpha}}

\newcommand{\Jc}{J_{\rm c}}

\newcommand{\Ji}{J_{\rm i}}

\newcommand{\ddt}[1]{\frac{\partial #1}{\partial t}}
\newcommand{\ddnu}[1]{\frac{\partial #1}{\partial \nu}}

\newcommand{\ddz}[1]{\frac{d #1}{dz}}

\newcommand{\nh}{n_{\rm H}}
\newcommand{\fhe}{f_{\rm He}}
\newcommand{\xe}{x_{\rm e}}

\newcommand{\xh}{x_{{\rm HI}}}
\newcommand{\mh}{m_{\rm H}}

\newcommand{\Tg}{T_{\gamma}}
\newcommand{\Tb}{T_{\rm b}}
\newcommand{\Tk}{T_{\rm k}}
\newcommand{\Ts}{T_{\rm s}}

\newcommand{\Dk}{D_{\rm k}}

\newcommand{\kb}{k_{\rm B}}
\newcommand{\tceff}{T_{\rm c}}

\newcommand{\vth}{\bt v_{\rm th}}

\newcommand{\scA}{\mathcal{A}}

\newcommand{\tgp}{\tau_{\rm GP}}

\newcommand{\xcmb}{x_{\rm CMB}}
\newcommand{\Ek}{\mathcal{E}_{\rm k}}
\newcommand{\Gk}{\Gamma_{\rm k}}
\newcommand{\Elyac}{\mathcal{E}_{\lyam, \rm c}}
\newcommand{\Elyai}{\mathcal{E}_{\lyam, \rm i}}

\newcommand{\Ecmb}{\mathcal{E}_{\rm CMB}}
\newcommand{\Ecomp}{\mathcal{E}_{\rm Comp}}

\newcommand{\Gcmb}{\Gamma_{\rm CMB}}
\newcommand{\xc}{x_{\rm c}}
\newcommand{\lcdm}{$\Lambda$CDM}
\newcommand{\fdm}{f_{\rm DM}}

\newcommand{\rhob}{\rho_{\rm b}}

\newcommand{\utht}{u_{{\rm th}, t}}
\newcommand{\me}{m_{\rm e}}
\newcommand{\se}{\sigma_{0, \rm e}}
\newcommand{\spn}{\sigma_{0, \rm p}}
\newcommand{\re}{r_{\rm e}}
\newcommand{\rp}{r_{\rm p}}

\newcommand{\Vb}{{\bt V}_{\rm b}}
\newcommand{\Vchi}{{\bt V}_{\chi}}
\newcommand{\Vchib}{{\bt V}_{\chi \rm b}}

\newcommand{\Edms}{\mathcal{E}_{\rm DM, s}}
\newcommand{\Edma}{\mathcal{E}_{\rm DM, a}}

\newcommand{\mnras}{Monthly Notices of the Royal Astronomical Society}
\newcommand{\apjl}{Astrophys. J. Lett.}
\newcommand{\aj}{Astronomical Journal}
\newcommand{\aap}{Astronomy and Astrophysics}

\newcommand{\sk}[1]{}

\begin{document}

\title{Heating of the intergalactic medium by the cosmic microwave background during cosmic dawn}
\author{Tejaswi Venumadhav}
\email{tejaswi@ias.edu}
\author{Liang Dai\footnote{NASA Einstein Fellow}}
\author{Alexander Kaurov}
\author{Matias Zaldarriaga}
\affiliation{Institute for Advanced Study, 1 Einstein Drive, Princeton, NJ 08540, USA}

\date{\today}

\begin{abstract}
The intergalactic medium is expected to be at its coldest point before the formation of the first stars in the universe. 
Motivated by recent results from the EDGES experiment, we revisit the standard calculation of the kinetic temperature of the neutral gas through this period. 
When the first ultraviolet (UV) sources turn on, photons redshift into the Lyman lines of neutral hydrogen and repeatedly scatter within the \lya line. 
They heat the gas via atomic recoils, and, through the Wouthuysen-Field effect, set the spin temperature of the 21--cm hyperfine (spin-flip) line of atomic hydrogen in competition with the resonant cosmic microwave background (CMB) photons. 
We show that the \lya photons also mediate energy transfer between the CMB photons and the thermal motions of the hydrogen atoms.
In the absence of X-ray heating, this new mechanism is the major correction to the temperature of the adiabatically cooling gas ($\sim 10 \%$ at $z=17$), and is several times the size of the heating rate found in previous calculations. 
We also find that the effect is more dramatic in non-standard scenarios that either enhance the radio background above the CMB or invoke new physics to cool the gas in order to explain the EDGES results.
The coupling with the radio background can reduce the depth of the 21--cm absorption feature by almost a factor of two relative to the case with no sources of heating, and prevent the feature from developing a flattened bottom. 
As an inevitable consequence of the UV background that generates the absorption feature, this heating should be accounted for in any theoretical prediction.

\end{abstract}

\maketitle

\section{Introduction}
\label{sec:introduction}

Within the standard Lambda cold dark matter (\lcdm) cosmology, the thermal history of the baryons can be accurately predicted until the onset of the first star formation. 
The baryons kinetically decouple from the cosmic microwave background (CMB) radiation at the end of cosmological recombination (at redshift $z_{\rm rec} \approx 1100$, at which point they are at a temperature of $\approx 3000$ K). 
Compton scattering off the residual free electrons keeps baryons thermally coupled to the CMB until redshifts of $z \approx 150$. 
Subsequently, the neutral gas adiabatically cools as the universe expands; it is expected to reach a temperature of $9.4$ K at $z = 20$ ($6.9$ K at  $z = 17$). 
Both the epoch of Compton decoupling, and the gas temperatures achieved at later times, depend sensitively on the residual free electron fraction post recombination. 
The process of cosmological recombination has been calculated to $\lesssim 0.1 \%$ level precision \cite{2011MNRAS.412..748C,2011PhRvD..83d3513A}.

The subsequent evolution of the baryonic temperature (at $z \lesssim 20$) is believed to have been driven by less well-constrained astrophysical sources. 
The ultraviolet (UV) radiation from the first stars resonantly scatters with the hydrogen atoms as it redshifts into the \lya line (as well as other Lyman-series lines) in the expanding Universe, and moderately heats the gas through atomic recoils \cite{2004ApJ...602....1C}. The first sources of X-rays heat the gas by photo-ionizing it and producing secondaries \cite{1997ApJ...475..429M,2010MNRAS.404.1869F}. 

Resonant scattering of \lya photons  also pumps the spin-flip (21--cm) transition between the hyperfine sub-levels of the ground $1s$ state---the so-called Wouthuysen-Field (WF) effect \cite{1952AJ.....57R..31W,1958PIRE...46..240F}. 
This process drives the spin temperature of the transition downward from the CMB temperature (to which it was previously set, due to magnetic dipole interactions with resonant 21--cm photons) toward the gas temperature \cite{2004ApJ...602....1C,2006MNRAS.367..259H}. 
Radiative transfer of the CMB photons through the gas produces a characteristic absorption feature, which effectively probes the UV radiation background and the temperature of the intergalactic medium (IGM) \cite{2005ApJ...626....1B,2006ApJ...637L...1K,2006ApJ...646..681S}. 
When integrated over a range of redshifts, this feature leaves a signature in the global radio background~\cite{1999A&A...345..380S}.

Recently, the EDGES experiment reported the most precise measurement to date of the global background in the $50-90$ MHz range~\cite{2018Natur.555...67B}. 
After carefully modeling their instrumental response in the presence of much larger foregrounds, they inferred an absorption feature of amplitude $0.5$ K at a frequency of $78$ MHz, which corresponds to a redshift of $z = 17$ for the 21--cm line. 
The amplitude and flattened shape of the reported absorption dip are larger than the expected maximum value of $0.22$ K (achieved if the spin temperature equals the gas temperature), and inconsistent with models of the early X-ray heating~\cite{2017MNRAS.472.1915C}. 

These results have stimulated renewed interest in the thermal state of the neutral IGM at redshifts $z \lesssim 17$. 
The large absorption feature has been interpreted as a sign of excess cooling of the gas due to baryon--dark matter (DM) interactions \cite{2018Natur.555...71B}, or of the presence of an excess radio background above the CMB~\cite{2018arXiv180207432F,2018arXiv180301815E}. 
The flattened shape of the dip has been interpreted to be due to a lack of X-ray heating~\cite{2018Natur.555...67B}.

In this paper, we show that when the spin temperature of the 21--cm line departs from the brightness temperature of the radio background, a new source of heating is activated: resonant radio photons in the vicinity of the 21--cm line indirectly transfer energy to the random motions of the gas through the energy reservoir associated with the spin states. 
This transfer occurs through the \lya photons; thus, the \lya photons not only directly supply energy, but also act as a conduit between the radio background and the gas. 
We show that the resultant heating rate is substantially higher than the rate solely due to the \lya photons. 

In standard \lcdm \, cosmology, the heating due to the CMB leads to a small correction to the thermal history and 21--cm absorption feature at cosmic dawn. 
We consider a few examples of non-standard thermal histories (such as the ones invoked to explain the EDGES results) and show that the new heating significantly impacts the temperature and  absorption feature, even in the absence of X-ray heating. 
Hence, any attempt to explain the size and shape of a putative global signal should account for the additional heating that we compute. 

The rest of this paper is organized as follows: Section \ref{sec:simple} provides a simple intuitive explanation of the new heating mechanism. 
Section \ref{sec:heatingcalc} contains the detailed calculation of the effect, including all the relevant microphysical processes. 
Section \ref{sec:results} estimates the size of the effect, both in standard and nonstandard thermal histories. 
We collect some technical details about baryon-DM scattering into Appendix~\ref{sec:bdm}. 
Finally, Section \ref{sec:conclusion} contains our conclusions. 

We use the Planck 2015 cosmological parameters \cite{2016A&A...594A..13P} as implemented in the {\sf Astropy} package~\cite{2018arXiv180102634T}, and, where applicable, the standard thermal and ionization histories computed using the {\sf HyRec} code~\cite{2011PhRvD..83d3513A}. 
In the rest of the paper, the phrase `CMB heating' refers to heating by any radio background in the vicinity of the 21--cm line.

\section{Simple explanation of the mechanism} 
\label{sec:simple}

The following energy reservoirs are active in the high redshift IGM: the thermal motions of the hydrogen atoms ($\vth$), their spins in the $1s$ state, the CMB photons, and the \lya photons (we ignore X-rays in our calculation). 
For the purposes of calculating the heat transfer, these reservoirs are at the kinetic ($\Tk$), spin ($\Ts$), and CMB ($\Tg$) temperatures, and the effective \lya color temperature ($\tceff$), respectively.\footnote{Our usage of `temperature' is an abuse of terminology, since in thermodynamics, the temperature of a heat bath is defined by the canonical ensemble according to which its internal degrees of freedom are populated. 
The CMB temperature $\Tg$ and the \lya color temperature $\tceff$ are defined from the flux on the blue side of the 21--cm line (via the Rayleigh-Jeans law) and the spin flip rate due to \lya scattering, respectively (both radiation fields need not be thermal).}

\begin{figure}[t]
  \centering
  \begin{tikzpicture}[>=stealth]
    \node[draw, text width=1cm, align=center] at (-2, 4) (CMB) {CMB \\ ($\Tg$)};
    \node[draw, text width=1.2cm, align=center] at (0, 2) (Sp) {HI spins \\ ($\Ts$)};
    \node[draw, text width=1cm, align=center] at (4, 2) (Lya) {$\lyam$ \\ ($\tceff$)};
    \node[draw, text width=1.2cm, align=center] at (0, -2) (Kin) {HI $\vth$ \\ ($\Tk$)};
    \node[draw, circle, align=center] at (-1, 3) (rad) {1};
    \node[draw, circle, align=center] at (0, 0) (coll) {2};
    \node[draw, circle, align=center] at (2, 0) (scatt) {3};
    \node (CMBd) [below of=CMB, node distance=1.5cm] {};
    \node (Kinu) [above of=Kin, node distance=1cm] {};
    
    \path[black, thick, ->] (CMB) edge (rad);
    \path[black, thick, ->] (rad) edge (Sp);
    \path[black, thick, ->] (Lya) edge node [below, xshift=0.25cm] {(a)} (scatt);
    \path[black, thick, ->] (Sp) edge node [above, xshift=0.25cm] {(b)}  (scatt);
    \path[black, thick, ->] (scatt) edge node [below, xshift=0.25cm] {(c)} (Kin);
    \path[black, dashed, ->] (Sp) edge (coll);
    \path[black, dashed, ->] (coll) edge (Kin);

    \draw[red, thick, ->, out=-45, in=45, distance=1.7cm] (CMBd) to (Kinu);

    \node[align=left] at (3.5, 4.5) (Dip) {1. Magnetic dipole transitions};
    \node[align=left] at (2.67, 4.0) (Coll) {2. HI--HI collisions};
    \node[align=left] at (2.95, 3.5) (WF) {3. \lya scattering};
  \end{tikzpicture}
  \caption{\label{fig:schematic} Schematic illustration of heat transfer during cosmic dawn: squares show energy reservoirs (along with their temperatures) that are coupled by microphysical processes (indexed by numbers). Arrows show the direction of heat transfer. The dashed line indicates that atomic collisions are inefficient. The red line is our new heating mechanism. Symbols (a), (b), and (c) label the net diffusive heat loss of the \lya photons, and the contributions of spin-flips and atomic recoils, respectively.}
\end{figure}
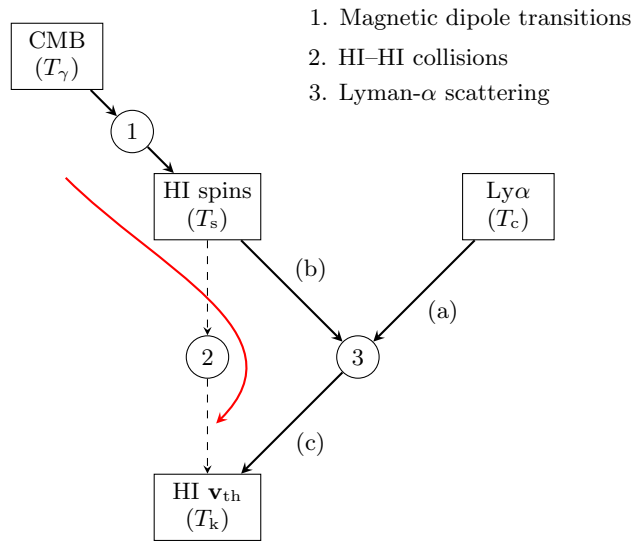

Figure \ref{fig:schematic} schematically shows these energy reservoirs along with the microphysical processes responsible for coupling them. 
The magnetic dipole transitions between the hyperfine levels due to the CMB photons drive the spin temperature $\Ts$ toward the CMB temperature $\Tg$. 
Spin flips due to collisions between neutral atoms tend to drive $\Ts$ toward the kinetic temperature $\Tk$ \cite{1969ApJ...158..423A,2005ApJ...622.1356Z}. 
Finally, UV photons within the \lya line resonantly scatter off the neutral hydrogen atoms, and in the process, cause two-step transitions between the hyperfine levels. 
This process couples three reservoirs together: when \lya photons scatter off neutral hydrogen atoms, they exchange energy both with the thermal motions ($\vth$) due to atomic recoil, as well as with the spins due to the WF effect (these are the kinetic and spin diffusivities of Ref.~\cite{2006MNRAS.367..259H}).
As a result, \lya scattering tends to drive $\Ts$, $\Tk$, and the \lya color temperature $\tceff$ toward a common value.\footnote{Collisions of neutral hydrogen atoms with free electrons and protons can also cause spin flips. 
  These are important only at ionization fractions $\xe \gtrsim 0.01$, while at $z = 20$, $\xe = 2 \times 10^{-4}$~\cite{2007MNRAS.374..547F,2007MNRAS.379..130F}. 
  Repeated scattering within the 21--cm line can also couple three reservoirs together, similarly to \lya scattering. 
  This process has several novel cosmological applications~\cite{2017PhRvD..95h3010V,2017arXiv170703513H}, but it is subdominant due to the low 21--cm optical depth.}

The standard calculation of the spin temperature $\Ts$ balances the rate of spin flips due to all the above processes (i.e., the rates of all arms that connect to the HI spins in Fig.~\ref{fig:schematic}. See, e.g., Ref.~\cite{2012RPPh...75h6901P}). 
From a thermodynamic perspective, the HI spins are a poor energy reservoir that immediately transfers the energy given by the CMB photons to the \lya photons, who in turn share it with the thermal motions of the atoms (neglecting collisions; if collisions are active, the energy is partitioned between arms 2 and (b) of Fig.~\ref{fig:schematic}). 

How much energy can this mechanism transfer to the gas? 
The blackbody of the CMB has a large amount of energy, but only photons within the 21--cm line cause spin flips. 
If we consider only the portion of the CMB that redshifts through the 21--cm line in a Hubble time, the energy content (in units of temperature per hydrogen atom) is
\begin{align}
   \frac{\Delta E_{\rm CMB}}{(3/2)\,\nh\, \kb} & = \frac{16\, \pi\, \nu_{21}^3}{3\, \nh\, c^3}\, T_\gamma = 60 \, {\rm K} \times \left( \frac{1+z}{20} \right)^{-2},
  \label{eq:Ecmb}
\end{align}
where $\nu_{21} = 1.4$ GHz and $\nh$ is the hydrogen number density. 
Only a fraction of order the 21--cm optical depth $\tau_{21}$ of this energy is absorbed, where 
\begin{align}
  \tau_{21} & = \frac{3}{32\pi} \frac{c^3 \nh \xh A_{21}}{\nu_{21}^3 H} \frac{T_{21}}{\Ts} \label{eq:tau21} \\
  & \approx 8.1 \times 10^{-2} \times \xh \left( \frac{1+z}{20} \right)^{3/2} \left( \frac{10 \, {\rm K}}{\Ts} \right).
\end{align}
Here $A_{21} = 2.86 \times 10^{-15} \, {\rm s}^{-1}$ is the Einstein A-coefficient of the 21--cm transition. 
This equation neglects slowly varying factors of $\nu/\nu_{21}$, and assumes that $\Ts \gg T_{21} = h \nu_{21}/\kb = 68.2 \, {\rm mK}$. 
The optical depth is small, but it is raised by any mechanism that additionally cools the gas. 

\section{Calculation of the heating rate}
\label{sec:heatingcalc}

In this section, we present a detailed calculation of the heating rate due to the CMB photons. 
We first revisit previous calculations of \lya heating and identify the corrections that are needed to incorporate the CMB heating, which will give us additional physical insight into the mechanism. 
We then present a simple alternative approach, and derive a final equation for the evolution of the temperature. 

The heating due to \lya scattering depends on the shape of the spectrum distortion in the vicinity of the \lya line.
If we can locally approximate the redistribution of photons within the line as a random walk in frequency, and assume that scattering is fast relative to the Hubble rate, we can obtain the \lya spectral distortion as the steady state solution of the Fokker-Planck equation \cite{1994ApJ...427..603R}.

According to the arguments in Ref.~\cite{2004ApJ...602....1C} (henceforth CJ04), the continuous energy loss of the photons to the baryons keeps the \lya distortion at a fixed physical frequency (instead of simply redshifting away due to the Hubble expansion).
A detailed balance argument relates the net rate of energy loss to the area under the spectral distortion.
This yields the rate of energy transfer in arm (a) of Fig.~\ref{fig:schematic}. 
CJ04 ignore the spin diffusivity and associated frequency drift (i.e., the up- and down-scattering of \lya photons due to spin flips), and hence do not have arm (b) of Fig.~\ref{fig:schematic} in their calculation of the \lya spectral distortion. 
Under this approximation, the energy flow to the atoms in arm (c) equals the energy loss from the \lya photons in arm (a).

However, when the spin and CMB temperatures satisfy $\Ts < \Tg$, the CMB systematically loses energy to the spins, and thus it is not consistent to ignore the energy transfer in arm (b) (which equals that in arm 1).
The resolution of the paradox is that without spin diffusivity, the color temperature $\tceff \approx \Tk$ due to the large \lya optical depth, and all heating comes from the wings of the line. 
When we include spin diffusivity and impose that $\Tg > \Tk$, we inevitably have $\Tg > \Ts > \tceff > \Tk$. The additional heating that arises after the CMB is included into the consideration of energy conservation comes from within the Doppler core of the \lya line.

We now show this directly from the \lya spectral distortion. 
We separately solve for the distortions sourced by continuum and injected fluxes, which, respectively, describe photons that directly redshift into the \lya line, and photons that redshift into higher Lyman-series lines and get reprocessed into \lya through radiative cascades. 
Within a small frequency range around the \lya line, the steady-state Fokker-Planck equation takes the form
\begin{align}
    -\scA(\nu) J(\nu) + D(\nu) \ddnu{J(\nu)} & = 
    H\,\nulya
    \begin{cases} 
      \Jc  \\
      \Ji\,\int_{\nu}^{\infty} d\nu^\prime \, \psi(\nu^\prime)
    \end{cases}.
  \label{eq:fp}
\end{align}
Here, $J(\nu)$ is the specific photon-number flux (hereafter flux) at frequency $\nu$ in units of ${\rm cm}^{-2} {\rm s}^{-1} {\rm Hz}^{-1} {\rm Sr}^{-1}$,  $\scA(\nu)$ and $D(\nu)$, respectively, are the frequency drift (including Hubble drift) and diffusivity in units of ${\rm Hz} \, {\rm s}^{-1}$ and ${\rm Hz}^2 \, {\rm s}^{-1}$, $H$ is the Hubble rate at the redshift of interest, $\nulya = 2466 \, {\rm THz}$ is the \lya frequency, $J_{\rm c, i}$ are input continuum and injected fluxes, and $\psi(\nu)$ is the probability with which photons are injected at frequency $\nu$.

We use kinetic and spin contributions to $\scA(\nu)$ and $D(\nu)$ that incorporate the fine and hyperfine splitting of the \lya line, provided in Ref.~\cite{2006MNRAS.367..259H} (henceforth H06). 
For the injection profile $\psi(\nu)$, we assume that cascades populate the four hyperfine sublevels of $2p$ according to their statistical weight. 
We discretize Eq.~\eqref{eq:fp} and solve it using a matrix method.\footnote{Numerical integration of \eqref{eq:fp} is stable only when done from the red side of the line. 
The right boundary condition to apply is $J(-\infty) = J_{\rm c, i}$, which can be seen by noting that $D(\nu)$ (and the scattering contribution to $\scA(\nu)$) vanish far from the \lya line, where $\scA(\nu) = - H \nulya$ (see H06).}

The next step is to compute heating rates from the spectral distortions. We work with the dimensionless heating efficiency $\Ek$, defined separately for the continuum and injected cases:
\begin{align}
  \Ek & = \frac{\Gk}{(3/2) \nh H \kb \Tk} \biggr\vert_{J_{\rm c, i} = J_0}, \, {\rm where} \label{eq:heff} \\
  J_0 & = \frac{\nh c}{4 \pi \nulya}. \label{eq:oneptperh}
\end{align}
Here, $\Gk$ is the volumetric heating rate in arm (c) of Fig.~\ref{fig:schematic}, and the flux scale $J_0$ corresponds to one photon per hydrogen atom.

The heating rate $\Gk$ due to any process can be written as an integral over the derivative of the line profile weighted by the corresponding diffusivity (see, e.g., Ref.~\cite{2013ApJ...779..146O}). 
Specializing to atomic recoil, we have
\begin{align}
  \Ek
  & = - \frac{8 \pi}{3 \nh c H \kb \Tk} \times \notag \\
  & ~~~\int d\nu \, h\nu \frac{\partial}{\partial \nu} \left[ \Dk(\nu) \left\{ \ddnu{J(\nu)} + \frac{h J(\nu)}{\kb \Tk} \right\} \right], \label{eq:hefffp}
\end{align}
where $\Dk(\nu)$ is the kinetic diffusivity.  
 
From the form of the diffusivities in H06, both the solution of the Fokker-Planck equation and the heating efficiency depend only on three parameters: the kinetic and spin temperatures, and the Gunn-Peterson optical depth 
\begin{align}
  \tgp & = \frac{3\,c^3\,\nh\,\xh\,\gamma}{2\,H\,\nulya^3} \label{eq:tgp} \\
  & \approx 2 \times 10^6 \left( \frac{1+z}{20} \right)^{3/2}\,\xh,
\end{align} 
where we have assumed matter domination. $\xh$ is the neutral fraction, and $\gamma = 50 \, {\rm MHz}$ is the HWHM of the \lya line.

\begin{figure}[t]
  \centering
  \includegraphics[width=\columnwidth]{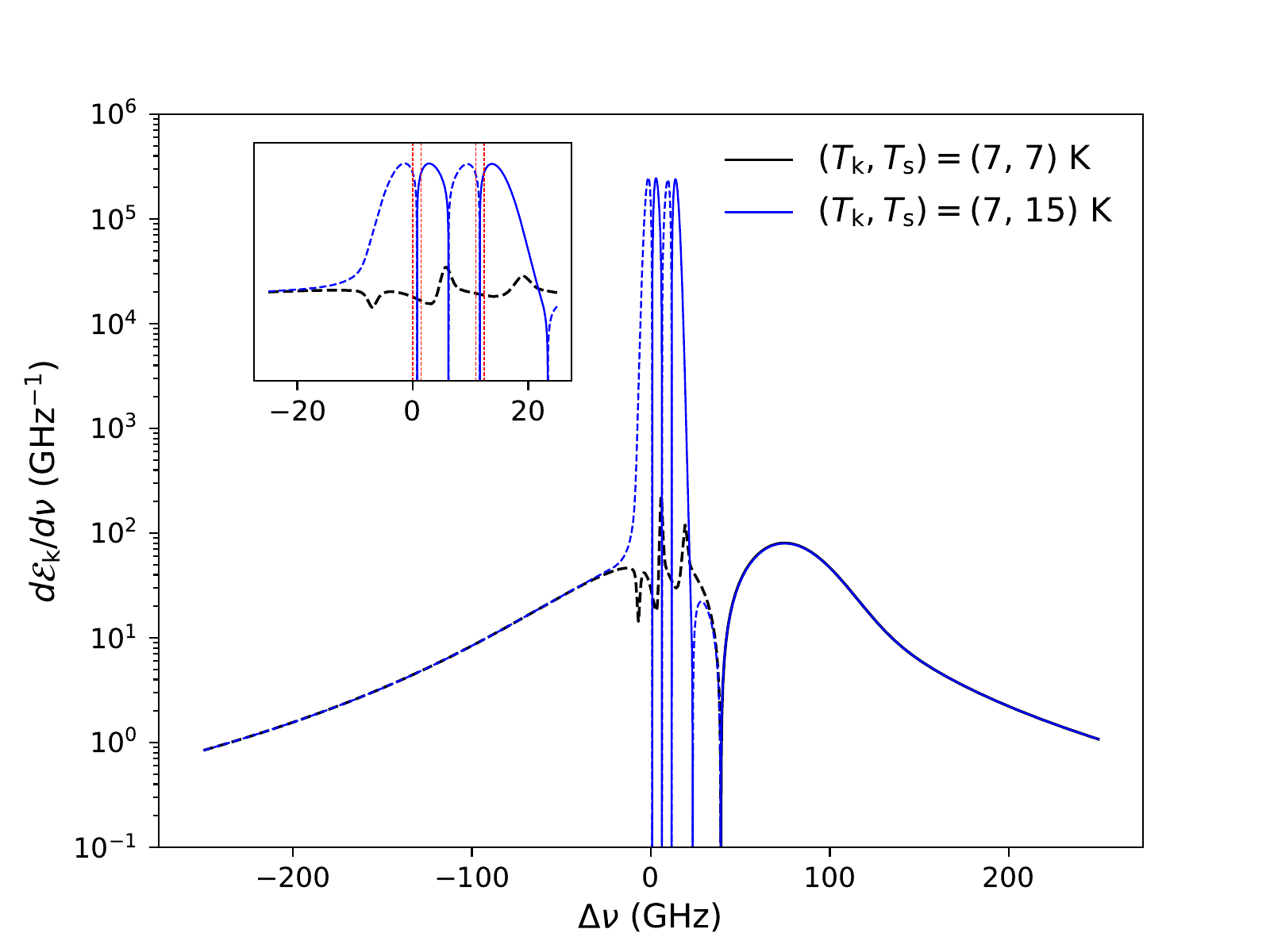}
  \caption{\label{fig:heatkernel} Heating within the \lya line: black and blue curves show the differential heating efficiency (see Eq.~\eqref{eq:hefffp}) vs frequency, for equal and unequal kinetic and spin temperatures, respectively (dashed parts correspond to negative values). In the first case, heating comes from the wings of the line, while in the second, the Doppler core dominates. The inset zooms into the core; vertical red lines mark the centers of the hyperfine components of the \lya line. The Gunn-Peterson optical depth $\tgp = 1.6 \times 10^6$.}
\end{figure}

Figure \ref{fig:heatkernel} shows the differential contribution to $\Ek$ vs frequency, computed using Eq.~\eqref{eq:hefffp}, for the case of continuum photons. 
It shows two cases: one with $(\Tk, \Ts) = (7, 7)$ K (such as would practically always happen if the WF effect were the only cause of spin flips), and one with $(\Tk, \Ts) = (7, 15)$ K, assuming $\tgp = 1.6 \times 10^6$. 
We see that in the former case, the heating is dominated by the extended wings of the \lya line (in agreement with CJ04), while in the latter, there are large positive and negative contributions within the Doppler core.

We cannot judge the net heating efficiency from Fig.~\ref{fig:heatkernel}. 
It is not advisable to directly integrate Eq.~\eqref{eq:hefffp}, since there are large cancellations between the heating and cooling effects of photons in different parts of the line. 
A numerically feasible procedure is to perform an integration by parts. 
Alternatively, we can add the contribution of arms (a) and 1 of Fig.~\ref{fig:schematic}: this cleanly separates the CMB contribution, and is valid even when atomic collisions are active during the dark ages (and the heat fluxes in arm (b) and 1 are not equal).

The first part is the pure \lya contribution (arm (a) of Fig.~\ref{fig:schematic}), which was calculated in CJ04. 
Following their approach, the heating efficiencies for continuum and injected photons (defined similarly to Eq.~\eqref{eq:heff}) are
\begin{subequations}
\label{eq:arm3eff}
\begin{align}
  \Elyac & = \frac{2 h}{3 \kb \Tk} \int d\nu \, \left[ 1 - \frac{J(\nu)}{\Jc} \right], \, {\rm and} \label{eq:arm3effc} \\
  \Elyai & = \frac{2 h}{3 \kb \Tk} \int d\nu \, \left[ \int_{\nu}^{\infty} d\nu^\prime \, \psi(\nu^\prime) - \frac{J(\nu)}{\Ji} \right]. \label{eq:arm3effi}
\end{align}
\end{subequations}
Using the above equations, we computed the values of $\Elyac$ and $\Elyai$ over a logarithmically spaced grid of spin and kinetic temperatures between $0.1 \, {\rm K}$ and $100 \, {\rm K}$, and Gunn-Peterson optical depths between $10^4$ and $10^7$. 
We use cubic spline interpolation to evaluate the efficiencies at general parameter values. 
The results are not expected to be accurate at the lowest temperatures because a) the assumption (made in the scattering line profiles) that $\Ts \gg T_{21}$ starts to break down, and b) the change in frequency during a scattering event can be larger than the width of the spectral feature, in which case the Fokker Planck equation does not apply.

The second part is the heating rate due to the CMB (arm 1 of Fig.~\ref{fig:schematic}). 
This rate equals the rate of spin flips multiplied by the energy transferred in each flip, $h\nu_{21} = 5.9 \, \mu \, {\rm eV}$. 
As noted in Sec.~\ref{sec:simple}, non-standard thermal histories with colder baryons can have larger 21--cm optical depths, in which case the spin-flip and heating rates should be computed accounting for the 21--cm spectral distortion itself.

If we assume that atomic processes are fast relative to the Hubble rate, the phase-space density $f(\nu)$ in the vicinity of the 21--cm line is
\begin{align}
  f(\nu) & = \frac{1}{T_{21}} \left[ \Ts + (\Tg - \Ts) e^{-\tau_{21} \, \chi_{21}(\nu)} \right], \label{eq:lineprofile}
\end{align}
where $\chi_{21}(\nu)$ is the cumulative function of the 21--cm line profile ($\chi_{21}$ goes from zero on the blue side to unity on the red side), $\Tg$ is the undistorted CMB temperature on the blue side of the 21cm line, and, as earlier, $T_{21} = h\nu_{21}/\kb = 68.2 \, {\rm mK}$. 
This equation is written under the same assumptions as Eq.~\eqref{eq:tau21} for the optical depth $\tau_{21}$.

\begin{figure*}[t]
  \centering
  \subfloat[][]{\includegraphics[width=\columnwidth]{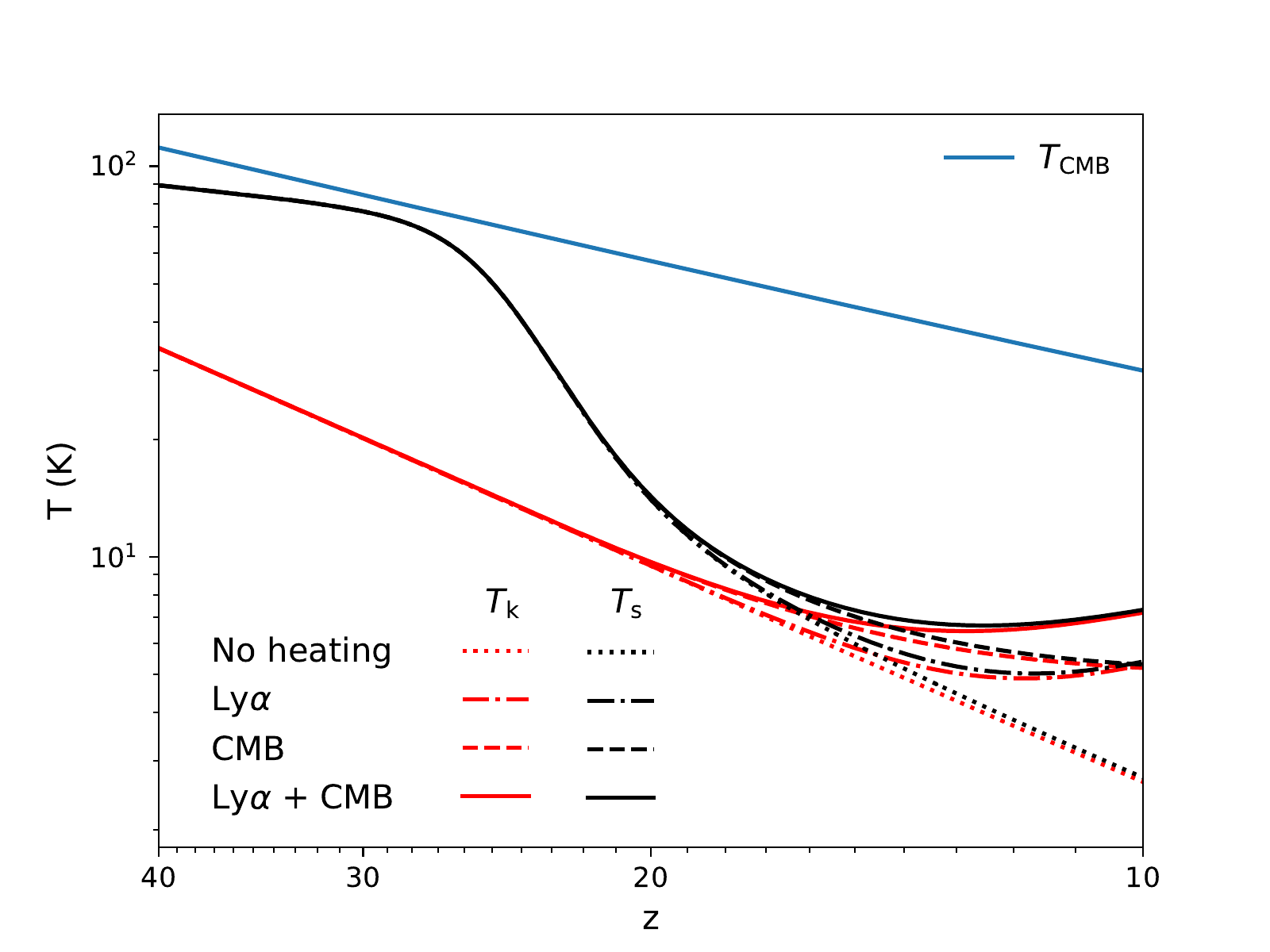} \label{fig:std_cmb_heating}}
  \subfloat[][]{\includegraphics[width=\columnwidth]{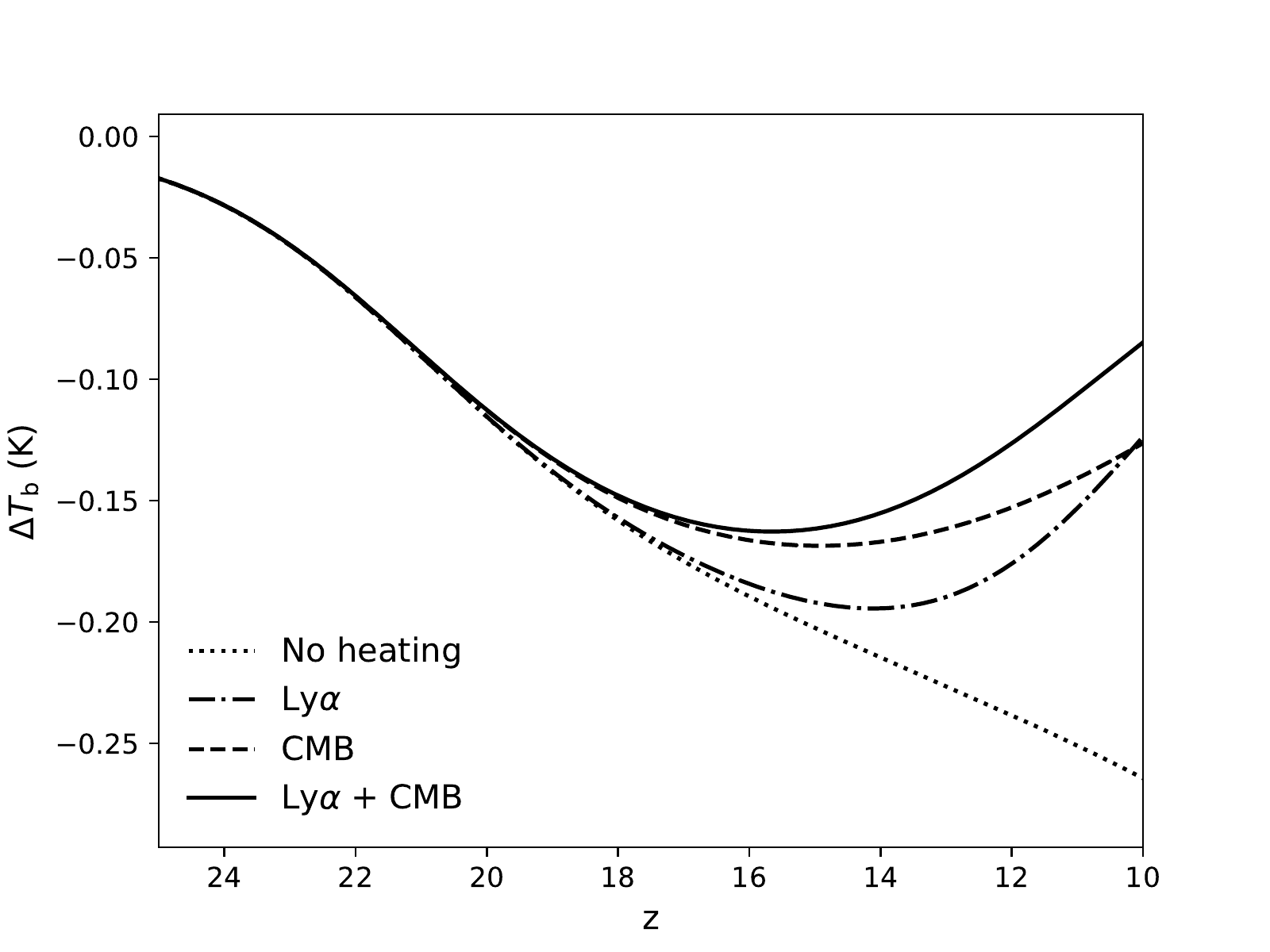} \label{fig:std_brightness_temp}}
  \caption{\label{fig:stdhistory} Temperatures vs redshift for standard parameters. All curves include the baseline Compton heating, which is negligible at these redshifts. {\em Left panel:} The red (black) dotted, dashed-dotted, dashed, and solid curves show the kinetic (spin) temperatures with no heating, only \lya heating, only CMB heating, and both \lya and CMB heating, respectively. {\em Right panel:} Differential brightness temperature against the CMB, $\Delta \Tb$.}
\end{figure*}

We obtain the rate of spin-flips due to the CMB by averaging over the line profile as follows:
\begin{align}
  \dot{y}_{\rm CMB} & = - A_{21} \left[ y - (3 - 4 y) \int_0^1 d\chi_{21}(\nu) \, f(\nu) \right], \label{eq:cmbspflip}
\end{align}
where $y \approx 3/4 - 3 T_*/16 \Ts$ is the triplet occupancy (see Ref.~\cite{2017PhRvD..95h3010V}; note the difference in the definition of $\chi_{21}$). 
The spin-flip rates due to the WF effect and atomic collisions are unchanged from those of H06. 
We substitute the line profile of Eq.~\eqref{eq:lineprofile} in Eq.~\eqref{eq:cmbspflip}, and balance the rate against those of all the other microphysical processes (all arms connecting to HI spins in Fig.~\ref{fig:schematic}). 
We get a modified version of the usual equation for the spin temperature:
\begin{align}
  \Ts^{-1} & = \frac{\xcmb \Tg^{-1} + \tilde{x}_\alpha \tceff^{-1} + \xc \Tk^{-1}}{\xcmb + \tilde{x}_\alpha + \xc}, \, {\rm where} \label{eq:tsinv} \\
  \xcmb & = \frac{1}{\tau_{21}} \left( 1 - e^{-\tau_{21}} \right),
\end{align}
and the coefficients $\tilde{x}_\alpha$ and $\xc$, respectively, represent the rate of spin-flips due to the WF-effect and atomic collisions relative to the rate due to the CMB (at low optical depth $\tau_{21}$). 
We use the rates in Ref.~\cite{2005ApJ...622.1356Z} to calculate the coefficient $\xc$. 
We numerically compute both $\tilde{x}_\alpha$ and the effective color temperature $\Tk$ from the \lya line profile, in the same manner as H06.
We have checked that our answers match the fitting formulae of H06 in the relevant range of parameters.

The brightness temperature $\Delta \Tb$ against the CMB is defined from the phase-space density on the red side of the line in Eq.~\eqref{eq:lineprofile}:
\begin{align}
  \Delta \Tb & = \xcmb \frac{\tau_{21}}{1 + z} \left( \Ts - \Tg \right). \label{eq:dtb}
\end{align}
We derive the heating efficiency due to the CMB from the spin-flip rate of Eq.~\eqref{eq:cmbspflip}:  
\begin{align}
  \Ecmb & = \frac{\Gcmb}{(3/2) \nh H \kb \Tk} \label{eq:heffcmb} \\
  & = \frac{\xh A_{21}}{2 H} \xcmb \left( \frac{\Tg}{\Ts} - 1 \right) \frac{T_{21}}{\Tk}, \label{eq:cmbheat}
\end{align}
where $\Gcmb$ is the volumetric heating rate. The heating rate in arm (b) of Fig.~\ref{fig:schematic} explicitly depends on the \lya flux, which is not immediately apparent in Eq.~\eqref{eq:cmbheat}. The dependence on the flux is encoded within the the spin temperature, through the coefficient $\tilde{x}_\alpha$ and the color temperature $\Ts$ in Eq.~\eqref{eq:tsinv}. 

The evolution of the kinetic temperature $\Tk$ depends on all the heating sources that are active. 
We add the heating efficiencies of \lya photons from Eq.~\eqref{eq:arm3eff}, the CMB from Eq.~\eqref{eq:cmbheat}, and add an additional Compton heating efficiency, to obtain the main result of our paper:
\begin{align}
  (1 + z) \ddz{\Tk} & = 2\,\Tk - \frac{1}{(1 + \fhe + \xe)} \times \notag \\
  &~~~ \left[ \Ecomp + \sum_{\rm r = \rm c, \rm i} \mathcal{E}_{\lyam, \rm r} \frac{J_{\rm r}}{J_0} + \Ecmb \right] \Tk, \label{eq:tevol}
\end{align}
where $J_{\rm c, i}$ are the continuum and injected \lya fluxes, $\fhe = 0.08$ is the helium-to-hydrogen number ratio, and $\xe = 1 - \xh$ is the ionization fraction. 
Note that all the quantities inside the brackets in the last term are functions of redshift.
Additionally, the Compton heating efficiency depends on the ionization fraction $\xe$. 
Ref.~\cite{2014PhRvD..89h3506A} derives the Compton heating rates and the evolution equation for $\xe$. 

\section{Results}
\label{sec:results}

\subsection{Standard temperature evolution}
\label{subsec:standard}

To judge the importance of CMB heating, we compute the temperature evolution for a toy \lya flux model. 
We compute spatially averaged \lya fluxes using the model in Ref.~\cite{2005ApJ...626....1B} and H06: we generate halos using the Sheth--Mo--Tormen mass function~\cite{2001MNRAS.323....1S}, and form Population III stars with an efficiency $f_\star = 0.01$ within those halos that can cool via atomic transitions ($T_{\rm vir} > 10^4$ K). 
We derive the continuum and injected \lya flux around halos assuming that the sources have blackbody spectra with temperatures of $10^5$ K. 

This model is simplistic in several respects: it does not include spatial variations in the UV background and in the kinetic temperature, assumes a constant and relatively large star formation efficiency within halos above a sharp mass threshold (which is at $2 \times 10^7 \, M_\odot$ at $z = 20$), and sets the baryon fraction within those halos to the cosmic mean value of $0.16$. 
Hence the results in this section should not be taken as predictions of the true values realized in the early universe, but rather as illustrations of the size of the CMB heating effect.

Figure \ref{fig:stdhistory} shows the evolution of temperatures for our toy \lya flux model, computed by integrating Eq.~\eqref{eq:tevol}. 
All the curves include the baseline Compton heating (the efficiency $\Ecomp$ in Eq.~\eqref{eq:tevol}), which is small at the redshift ranges shown. 
The dotted-dashed and dashed curves show results with only pure \lya heating (i.e., Eq.~\eqref{eq:arm3eff}) turned on and only CMB heating (i.e., Eq.~\eqref{eq:cmbheat}) turned on, respectively, while the solid curves show results with all the sources of heating turned on. 
Figure \ref{fig:std_cmb_heating} shows the kinetic ($\Tk$) and spin ($\Ts$) temperatures, while Fig.~\ref{fig:std_brightness_temp} shows the brightness temperatures against the CMB ($\Delta \Tb$) calculated using Eq.~\eqref{eq:dtb}. 

For our toy \lya flux model, and without any additional heating, the kinetic and brightness temperatures at $z = 17 \, (15)$ are $\Tk = 7 \, {\rm K}$ ($5.6 \, {\rm K}$) and $\Delta \Tb = -0.18 \, {\rm K}$ ($-0.2 \, {\rm K}$). 
The CMB heating alone makes a $\simeq 8.6 \%$ ($\simeq 15 \%$) correction to both $\Tk$ and $\Delta \Tb$ at $z = 17$ ($15$).
In comparison, \lya heating alone makes a $\simeq 1.3 \%$ ($\simeq 5 \%$) correction to both $\Tk$ and $\Delta \Tb$ at $z = 17$ ($15$).

\begin{figure*}[t]
  \centering
  \subfloat[][]{\includegraphics[width=\columnwidth]{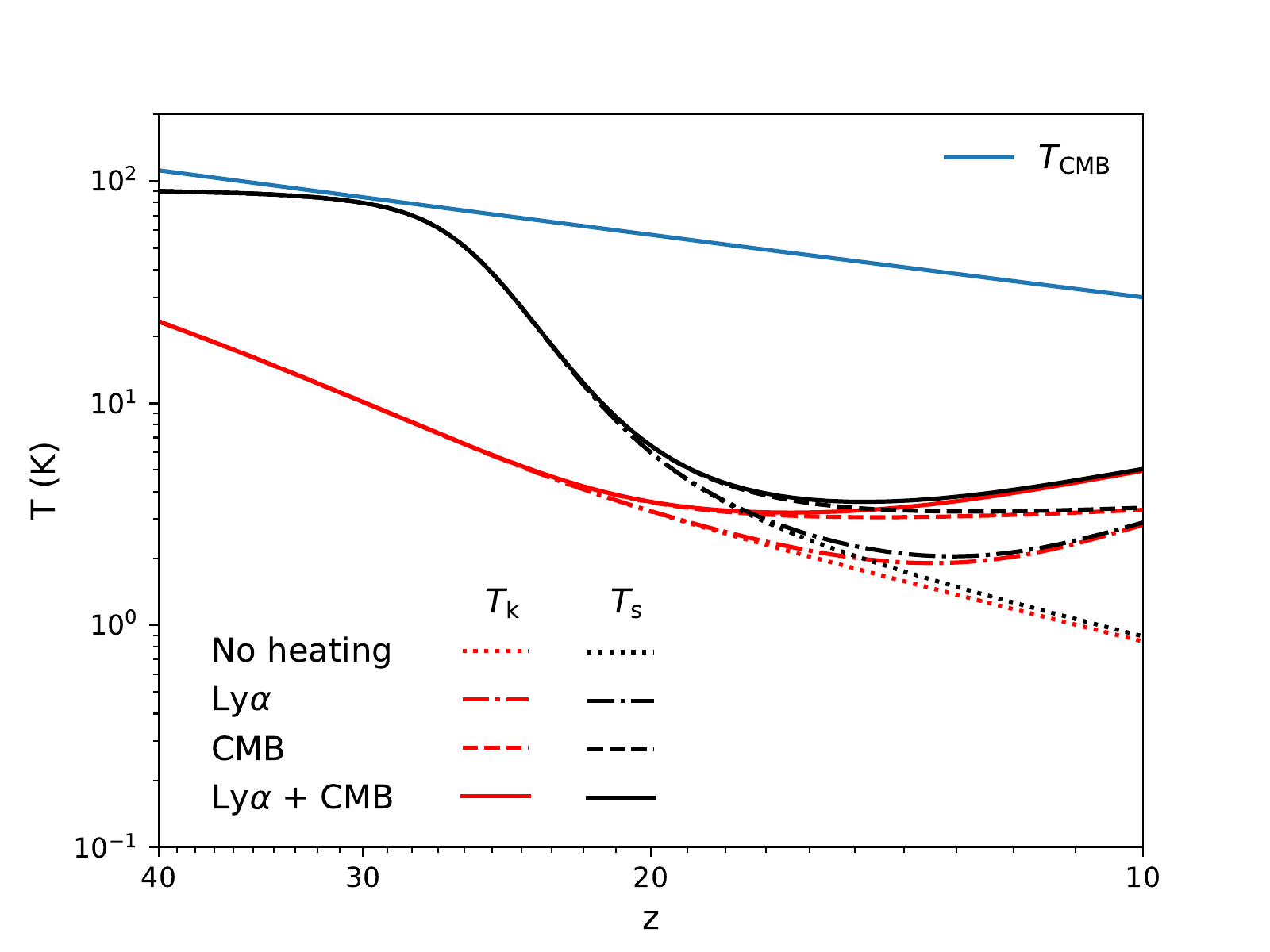} \label{fig:dm_cmb_heating}}
  \subfloat[][]{\includegraphics[width=\columnwidth]{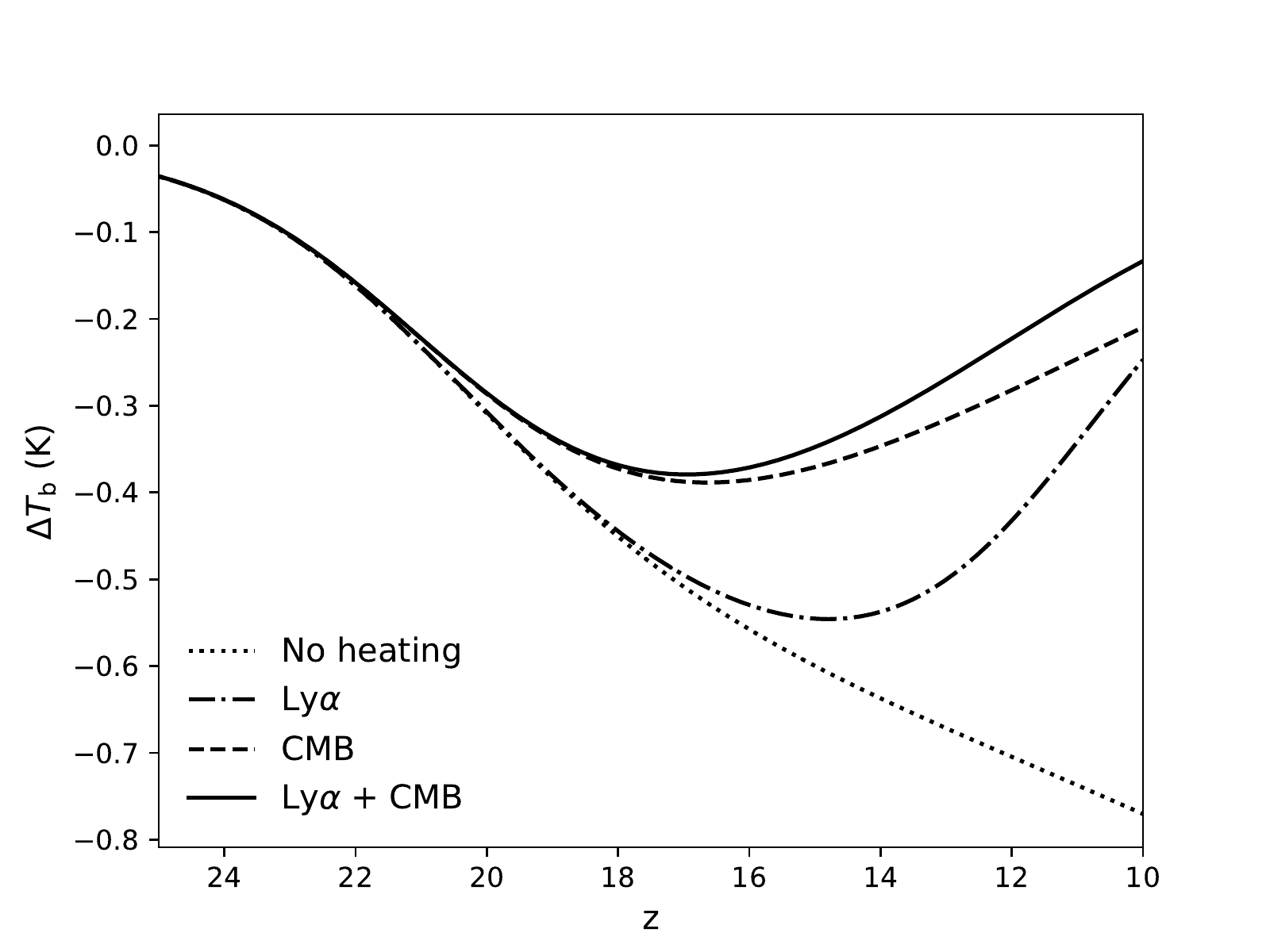} \label{fig:dm_brightness_temp}} \\
  \subfloat[][]{\includegraphics[width=\columnwidth]{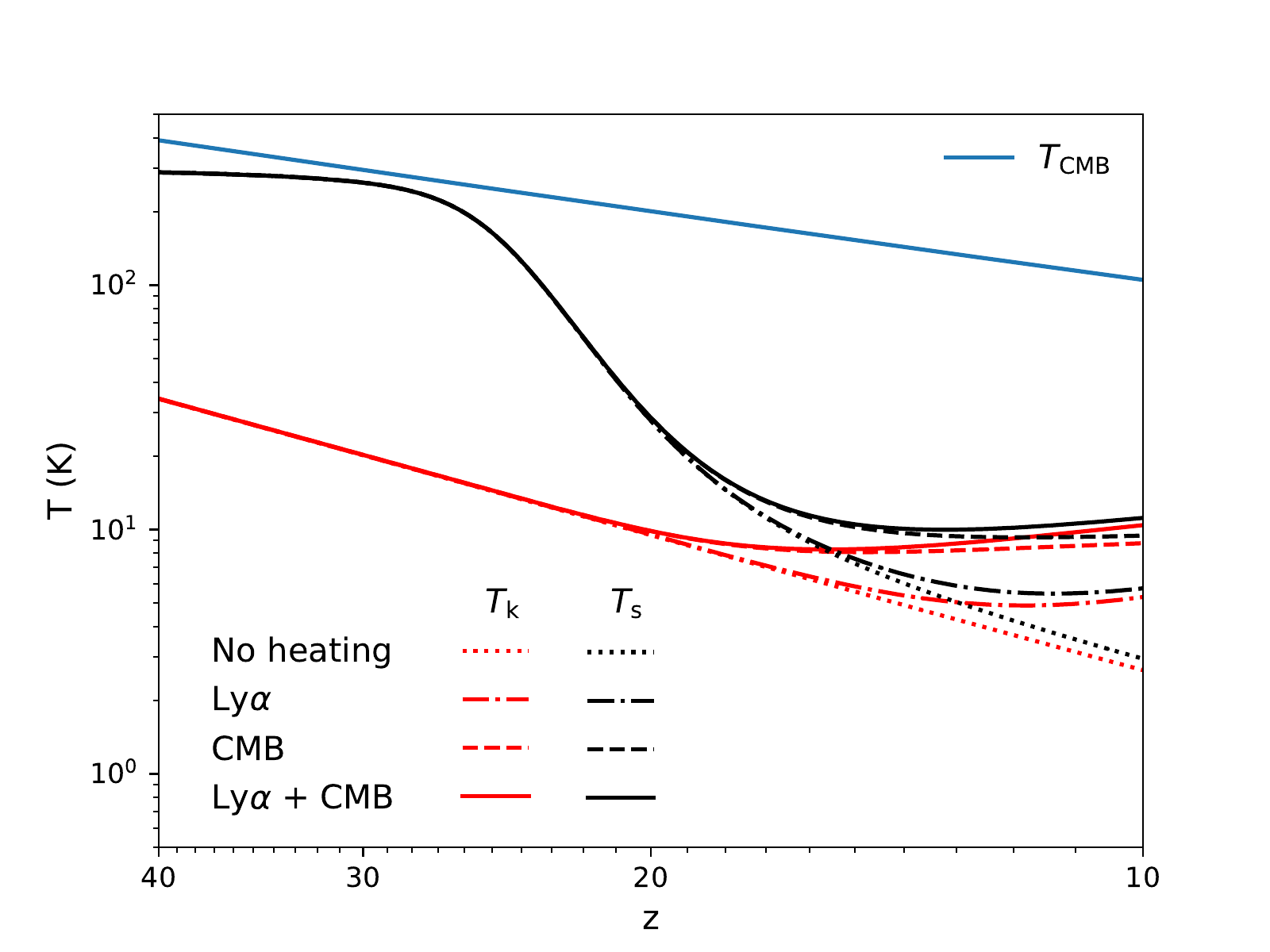} \label{fig:highcmb_cmb_heating}}
  \subfloat[][]{\includegraphics[width=\columnwidth]{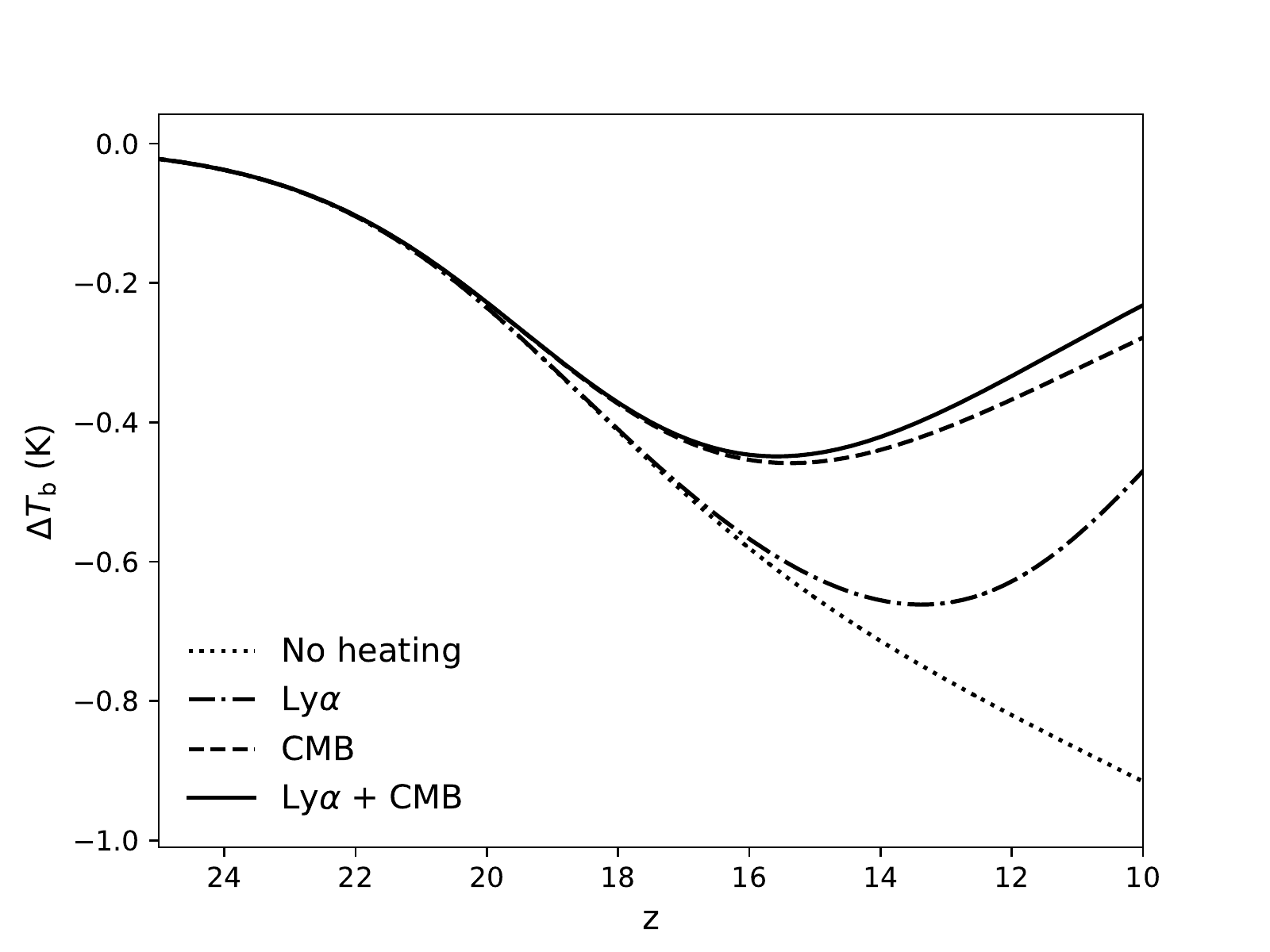} \label{fig:highcmb_brightness_temp}}
  \caption{\label{fig:nonstdhistory} Non-standard temperature evolution. Left and right panels in each row are as in Fig.~\ref{fig:stdhistory}. The modifications are as follows.  {\em Top row:} Baryon-dark matter ($\chi$) interactions with $m_\chi = 10 \, {\rm MeV}$, and a fraction $\fdm = 0.02$ of the DM having charge $\epsilon = 10^{-5}$, in a local patch with $\Vchib = 0$. {\em Lower row:} Radio background that boosts the CMB brightness temperature on the blue side of the 21--cm line, $\Tg$, by a factor of 3.5}
\end{figure*}

We have neglected X-ray heating, which is expected to be important after the formation of the first sources (see, e.g., Ref.~\cite{1997ApJ...475..429M} and CJ04). 
The amount of X-ray heating is highly uncertain; in several astrophysical scenarios considered in the literature (see, e.g., Ref.~\cite{2017MNRAS.472.1915C}), X-ray heating would dominate all the sources of heating considered in this paper. 
Hence the calculations in this paper should be important only in the absence of any significant X-ray heating.

\subsection{Non-standard temperature evolution}
\label{subsec:nonstandard}


Several recent papers have explored modifications to the standard temperature evolution to explain the large amplitude of the absorption dip reported by EDGES~\cite{2018Natur.555...71B,2018arXiv180210094M,2018arXiv180303091B,2018arXiv180302804B,2018arXiv180207432F,2018arXiv180301815E,2018arXiv180309739L}. 
In the limit of low values of the 21--cm optical depth $\tau_{21}$, the brightness temperature as given by Eq.~\eqref{eq:dtb} depends on the temperatures only through the combination $(1-\Tg/\Ts)$. 
In standard \lcdm \, cosmology, at this redshift, the temperatures satisfy $\Tg > \Ts > \Tk$. 
Modifications that increase the size of the dip do so by either reducing $\Tk$ (and hence $\Ts$) \cite{2018Natur.555...71B,2018arXiv180210094M,2018arXiv180302804B,2018arXiv180303091B,2018arXiv180309739L}, or increasing $\Tg$ \cite{2018arXiv180207432F,2018arXiv180301815E} (note that $\Tg$ here is the Rayleigh-Jeans brightness temperature on the blue side of the 21--cm line, and not necessarily the blackbody temperature of the CMB). 

We will illustrate the effect of CMB heating within both kinds of scenarios. 
In each case, we use the same model for the \lya flux as was used in Sec.~\ref{subsec:standard}. 
We use these examples to demonstrate the importance of CMB heating in non-standard thermal histories.

The first test case is the milli-charged dark matter model, in which a fraction of the DM has a small effective electric charge and interacts with the ionized part of the baryons. 
In this model, the interaction cross section between the DM and a charged target $t$ is $\sigma_t = \sigma_{0, t} (v_t/c)^{-4}$. 
This model is already severely constrained by a number of astrophysical and laboratory experiments; current estimates that include the EDGES results as a constraint (but do not include CMB heating, nor any models for the \lya flux) allow for a charged fraction of $\fdm \sim 1\%$ for DM particle masses in the range $10-80$ MeV~\cite{2018arXiv180210094M,2018arXiv180302804B,2018arXiv180303091B}.

The additional cooling depends on the bulk relative velocity between baryons and DM, $\Vchib$~\cite{2015PhRvD..92h3528M}. 
Appendix \ref{sec:bdm} presents our equations for the evolution of $\Vchib$, and the additional cooling efficiency due to the DM. 
As a concrete example, we consider the case where the charged DM particle is a Dirac fermion. 
In this case, we also include the extra heating due to the annihilation to standard model fermions \cite{2011PhRvD..83f3509M}. 

Figures \ref{fig:dm_cmb_heating} and \ref{fig:dm_brightness_temp} show the evolution of the spin and kinetic temperatures, and the 21--cm brightness temperature, respectively. 
The figure presents results within a patch with $\Vchib = 0$, for DM mass $m_\chi = 10 \, {\rm MeV}$, charge (in units of the electron charge) $\epsilon = 10^{-5}$, and charged fraction $\fdm = 0.02$. 
For these parameters, the characteristic cross sections $(\sigma_{0, \rm e}, \sigma_{0, \rm p}) = (3 \times 10^{-33}, 7.3 \times 10^{-36}) \, {\rm cm}^2$ (up to a logarithmic redshift dependence due to the cutoff in the momentum transfer).

After recombination, at the redshift of decoupling, the relative velocity between the bulk of the DM and the baryons is coherent on $\sim 100 \, {\rm Mpc}$ scales and has an RMS value of $29 \, {\rm km} \, {\rm s}^{-1}$~\cite{2010PhRvD..82h3520T}. 
However, given the masses and cross sections in Figs.~\ref{fig:dm_cmb_heating} and \ref{fig:dm_brightness_temp}, the charged fraction of the DM can be strongly coupled with the baryons prior to recombination~\cite{2018arXiv180309739L}. 
We checked that the relative velocity between these components is damped at higher redshifts, and that the answer is not extremely sensitive to the value of $\Vchib$ used.

Similar to Fig.~\ref{fig:stdhistory}, all curves include the baseline Compton heating. 
For these parameters, without any additional heating, the kinetic and brightness temperatures at $z = 17 \, (15)$ are $\Tk = 2.3 \, {\rm K}$ ($1.8 \, {\rm K}$) and $\Delta \Tb = -0.5 \, {\rm K}$ ($-0.6 \, {\rm K}$). 
The CMB heating alone makes a $\simeq 37 \%$ ($\simeq 70 \%$) correction to $\Tk$ and a $\simeq 24 \, \%$ ($\simeq 38 \, \%$) correction to $\Delta \Tb$ at $z = 17$ ($15$).
In contrast, the pure \lya heating alone makes a $\simeq 3 \%$ ($\simeq 11\%$) correction to $\Tk$ and a $\simeq 2.6 \%$ ($\simeq 9 \%$) correction to $\Delta \Tb$ at $z = 17$ ($15$).

The second class of modifications have an additional radio background above the CMB near the 21--cm, due to which the brightness temperature $\Tg > T_{\rm CMB} = 2.7255 \, {\rm K} \, (1+z)$ (an example is the extra radio background from black holes proposed in Ref.~\cite{2018arXiv180301815E}). 
These models are not easily described by a few parameters, so to illustrate the correction, we will consider a model in which the value of $\Tg$ is simply boosted by a constant factor.

Figures \ref{fig:highcmb_cmb_heating} and \ref{fig:highcmb_brightness_temp} show the results when $\Tg$ is 3.5 times its value in the standard \lcdm \, cosmology. 
In this case, without any additional heating, the kinetic and brightness temperatures at $z = 17 \, (15)$ are $\Tk = 7 \, {\rm K}$ ($5.6 \, {\rm K}$) and $\Delta \Tb = -0.5 \, {\rm K}$ ($-0.65 \, {\rm K}$). 
The CMB heating alone makes a $\simeq 19 \%$ ($\simeq 45 \%$) correction to $\Tk$ and a $\simeq 15 \%$ ($\simeq 30 \%$) correction to $\Delta \Tb$ at $z = 17$ ($15$).
This is to be compared to the correction of $\simeq 1.3 \%$ ($\simeq 5 \%$) to $\Tk$ and $\Delta \Tb$ at $z = 17$ ($15$) due to pure \lya heating alone.

In both the non-standard scenarios, the CMB heating dominates the pure \lya heating, and substantially suppresses the amplitude of the 21--cm absorption dip. 
As in Sec.~\ref{subsec:standard}, we have neglected any heating by early sources of X-rays. 
From Figs.~\ref{fig:dm_cmb_heating} and \ref{fig:highcmb_cmb_heating}, we also see that the effect of the CMB or the radio background is to heat the gas in a gradual manner. 
Even if the \lya flux were large enough to set $\Ts \approx \Tk$ and all X-ray heating were turned off (as has been assumed in all the calculations of the non-standard scenarios), CMB heating would produce an absorption feature whose slope varies smoothly with frequency.

\section{Conclusions}
\label{sec:conclusion}

Previous work has shown that \lya photons repeatedly scatter off neutral hydrogen and moderately heat the gas through atomic recoils. 
We have shown that HI spins and \lya photons act as mediators between the radio background and the random thermal motions of the HI atoms, and hence cause an extra heating of the IGM during cosmic dawn that has been neglected in previous calculations. 
Ultimately, this energy comes from the radio photons in the vicinity of the 21--cm transition line. 
The detailed heating rate depends on the difference between the spin and radio brightness temperatures, which in turn depends on the \lya flux. 
Within a simplified model of the UV background, and with the radio background given by the CMB, we estimate that the additional heating is several times more efficient than the pure \lya heating that was previously calculated. 

The CMB heating we have computed is an inevitable consequence of a UV background. 
It is the dominant physical process that modifies the thermal evolution of the IGM before the onset of X-ray heating from astrophysical sources. 
As such, this effect cannot be ignored in any explanation of the anomalously large absorption dip reported by the EDGES experiment. 

We use the same simplified UV background model as in the standard case to illustrate the size of the heating (rather than assuming that the \lya flux is extremely large, which is astrophysically hard to achieve), and consider two examples of non-standard temperature histories: a milli-charged dark matter scenario, and one with an extra radio background above the CMB in the wavelength range of the 21--cm transition. 
We find that in both scenarios, the new heating is dominant over the pure \lya heating, and substantially reduces the size of the absorption dip (at the factor of two level).

This heating can have two important consequences: firstly, models such as milli-charged dark matter have a very small parameter range that is viable in order to explain the EDGES results, and hence the constraints can be strongly affected~\cite{2018arXiv180302804B,2018arXiv180309739L}. 
Secondly, the flat-bottomed shape of the inferred absorption feature has been interpreted as the sign of delayed X-ray heating~\cite{2018Natur.555...67B}. 
The heating due to the radio background gradually changes the kinetic temperature of the gas, and thus tends to produce a smoother absorption feature.

We do not explore the quantitative changes to the DM constraints or to the shape of the absorption feature, since these necessarily require detailed modeling of the UV background and the fluctuations in the gas temperature (such as the work in Ref.~\cite{2018arXiv180303272M}) which is beyond the scope of this paper. 
Tabulated values of the \lya heating efficiencies (to be used in Eq.~\eqref{eq:tevol}), Wouthuysen-Field coupling coefficients, and color temperatures can be found at \url{https://github.com/ntveem/lyaheating}.

\begin{acknowledgments}

We thank Jordi Miralda-Escude, Julian Mu{\~n}oz, and Yacine Ali-Ha{\"\i}moud for reading an earlier version of the draft and sending us helpful comments.
TV and AK acknowledge support from the Schmidt Fellowship. TV is also supported by the W.M. Keck Foundation Fund. LD is supported at the Institute for Advanced Study by NASA through Einstein Postdoctoral Fellowship grant number PF5-160135 awarded by the Chandra X-ray Center, which is operated by the Smithsonian Astrophysical Observatory for NASA under contract NAS8-03060.

\end{acknowledgments}

\appendix

\section{Baryon-dark matter scattering}
\label{sec:bdm}

In this section, we write down the equations for the thermal evolution of a local patch of the universe in the presence of baryon-dark matter scattering. 
A detailed derivation of the energy exchanges, starting from the microphysics, was presented in Ref.~\cite{2015PhRvD..92h3528M} (henceforth, MKA15). 
However, they assumed that the DM scatters off all particles (i.e., even the neutral ones) and set the Helium fraction to zero. 

In what follows, the symbol $\chi$ labels the DM particle. Our equations are written under the following assumptions:
\begin{enumerate}
  \item The DM does not interact with the neutral component of the matter, i.e., its interaction with atomic dipole moments is negligible.
  \item The interactions between the charged components and the rest of the matter (and within the rest of the matter) are large enough to keep all baryonic components at a common velocity and temperature. 
  \item A fraction $\fdm < 1$ of the DM has a nonzero and small electric charge, and this part does not interact with the neutral fraction.
  \item The charged component is a Fermion with mass $m_\chi > 1 \, {\rm MeV}$, and the positively and negatively charged components have particle--anti-particle asymmetry.
\end{enumerate}
Our aim is not to survey all possible dark matter models, so we pick one model to illustrate the effect of the CMB heating. 

Charged DM particles scatter off ionized targets (indexed by $t$, which runs over electrons and protons) with a Rutherford-like cross section
\begin{align}
  \sigma_t & = \sigma_{0, t}\, (v_t/c)^{-4}, \, {\rm where} \\
  \sigma_{0, t} & = \frac{2\pi\,\alpha_{\rm EM}^2\,\epsilon^2\,\xi\,\hbar^2}{\mu_{t, \chi}^2\,c^2} \label{eq:sigma0} \\
  & = 5.7 \times 10^{-33} \, {\rm cm}^2 \, \left( \frac{\epsilon}{10^{-5}} \right)^2 \left( \frac{\mu_{t, \chi}}{0.5 \, {\rm MeV}} \right)^{-2} \left( \frac{\xi}{100} \right).
\end{align}
Here $\mu_{t, \chi} = m_\chi m_t/(m_\chi + m_t)$ is the reduced mass, $\epsilon$ is the charge of the DM in units of that of the electron, and $\xi$ is a factor that depends logarithmically on the cutoff in the momentum-transfer cross section~\cite{2011PhRvD..83f3509M,2018arXiv180210094M}. Note that the normalizations for different targets are related by $\sigma_{0, t} \propto \mu_{t, \chi}^{-2}$. Due to the small values of the DM charge, we can neglect the Compton scattering with the CMB photons. 

From the second assumption above, the equation of momentum conservation within a local patch is
\begin{align}
  \fdm\, \rho_\chi\, \ddt{\Vchi} \biggr\vert_{\rm s} + \rhob \ddt{\Vb} \biggr\vert_{\rm s} & = 0, \label{eq:pcons}
\end{align}
where $\rho_{\rm \chi, b}$ and ${\bt V}_{\chi, \rm b}$ are the mass density and bulk velocities of the DM and baryons, respectively (Note that $\Vchi$ is the velocity of the charged component of the DM). 
The subscript $s$ indicates that this is the contribution due to scattering. 
From Eq.~\eqref{eq:pcons}, the relative velocity between the charged DM and the baryons evolves according to
\begin{align}
  \ddt{\Vchib} \biggr\vert_{\rm s} & = \left( 1 + \fdm \frac{\rho_\chi}{\rhob} \right) \ddt{\Vchi} \biggr\vert_{\rm s}. \label{eq:dvchibdt}
\end{align}
MKA15 derive an expression for $\partial \Vchi/\partial t$. 
Specializing to the case where only a subset of the baryons scatters,
\begin{align}
  \ddt{\Vchi}\biggr\vert_{\rm s} & = -\int d^3 {\bt v}_\chi \, f_\chi({\bt v}_\chi) \sum_t \frac{\rho_t}{m_t + m_\chi} \times \notag \\
  & \hspace{20pt} \int d^3 {\bt v}_t \, f_t({\bt v}_t)  \, ({\bt v}_\chi - {\bt v}_t) \vert {\bt v}_\chi - {\bt v}_t \vert \bar{\sigma}_t, \label{eq:dvchidt}
\end{align}
where ${\bt v}_{\chi, t}$ are the microscopic velocities, the $f(\bt v)$s are the velocity distribution functions, and $\bar{\sigma}_t$ is the momentum transfer cross section. 
We substitute Eq.~\eqref{eq:dvchidt} in Eq.~\eqref{eq:dvchibdt}, and use the integrals over the velocity distributions computed in MKA15 to obtain
\begin{align}
  \ddt{\Vchib}\biggr\vert_{\rm s} & = - \left( 1 + \fdm \frac{\rho_\chi}{\rhob} \right) \frac{1}{V_{\chi \rm b}^2} \times \notag \\
  & \hspace{20pt} \sum_t \frac{\rho_t \sigma_{0, t}\,c^4}{m_t + m_\chi} F\left(r_t \equiv \frac{V_{\chi \rm b}}{\utht} \right), \, {\rm where} \\
  \utht^2 & = \frac{\kb\,\Tk}{m_{\rm t}} + \frac{\kb\,T_\chi}{m_\chi}, \, {\rm and} \\
  F(r) & = {\rm erf}\left( \frac{r}{\sqrt{2}} \right) - \sqrt{\frac{2}{\pi}} r \, e^{-r^2/2}.
\end{align}
Here $T_\chi$ is the temperature of the charged component of the dark matter, and from assumption 2 above, all the targets are at the common temperature $\Tk$. Letting the index $t$ run over electrons and protons, we have
\begin{align}
  \ddt{\Vchib}\biggr\vert_{\rm s} & = - \left( 1 + \fdm \frac{\rho_\chi}{\rhob} \right) \frac{c^4 \nh \xe}{V_{\chi \rm b}^2} \times \notag \\
  & \hspace{10pt} \left[ \frac{\me \se}{\me + m_\chi} F( \re ) + \frac{\mh \spn}{\mh + m_\chi} F( \rp ) \right].
\end{align}
The relative velocity evolves with redshift according to
\begin{align}
  (1 + z) \ddz{\Vchib} & = \Vchib - \frac{1}{H} \ddt{\Vchib}\biggr\vert_{\rm s}. \label{eq:dvchibdz}
\end{align}
We read off the heating rate per particle for each target `$t$' from the appendix of MKA15
\begin{align}
  \ddt{q_t} \biggr\vert_{\rm s} & = \fdm n_\chi \frac{m_t m_\chi \sigma_{0, t}\,c^4}{(m_\chi + m_t)^2 \utht^3} \times \notag \\
  & \hspace{20pt} \left[ \sqrt{ \frac{2}{\pi} } (\kb\,T_\chi - \kb\,\Tk) \, e^{-r_t^2/2} + m_\chi \utht^2 \frac{F(r_t)}{r_t} \right],
\end{align}
where $n_\chi = \rho_\chi/m_\chi$ is the number density of DM particles. The net heating rate per unit volume can be converted into an efficiency in exactly the same manner as in Eq.~\eqref{eq:heffcmb}:
\begin{align}
  \Edms & = \frac{1}{(3/2) \nh H \kb \Tk} \sum_{t} n_t \ddt{q_t} \\
  & = \frac23 \frac{\xe \fdm n_\chi}{H} \sum_{t = e, p} \frac{m_t m_\chi \sigma_{0, t} c^4}{(m_\chi + m_t)^2 \utht^3} \times \notag \\
  & \hspace{20pt} \left[ \sqrt{ \frac{2}{\pi} } \left( \frac{T_\chi}{\Tk} - 1 \right) \, e^{-r_t^2/2} + \frac{m_\chi \utht^2}{\kb \Tk} \frac{F(r_t)}{r_t} \right]. \label{eq:qbdot}
\end{align}
We have restored CGS units in the above equation. 

The heating rate of the charged component of the DM, in units of ${\rm erg} \, {\rm s}^{-1}$ per particle, equals
\begin{align}
  \ddt{q_\chi} & = \nh \xe \sum_{t=e, p} \frac{m_\chi m_t \sigma_{0, t} c^4}{(m_\chi + m_t)^2 \utht^3} \times \notag \\
  & \hspace{20pt} \left[ \sqrt{ \frac{2}{\pi} } \kb (\Tk - T_\chi) \, e^{-r_t^2/2} + m_t \utht^2 \frac{F(r_t)}{r_t} \right] \label{eq:qchidot}.
\end{align}
The DM temperature $T_\chi$ evolves with redshift according to
\begin{align}
  (1 + z) \ddz{T_\chi} & = 2\,T_\chi - \frac{1}{(3/2)\,\kb\,H} \ddt{q_\chi}. \label{eq:dtchidz}
\end{align}
Due to assumption 4 above, and for the relevant range of masses, the millicharged DM particles also annihilate into $e^\pm$. 
The velocity-averaged annihilation cross section is \cite{2011PhRvD..83f3509M}
\begin{align}
  \langle \sigma_{\rm a} v_{\rm rel} \rangle & = \sigma_{0, \rm e}\,c\, \frac{\mu_{\chi, \rm e}^2}{2 \xi m_\chi^2} \sqrt{1 - \frac{\me^2}{m_\chi^2}} \left( 1 + \frac{\me^2}{2 m_\chi^2} \right). \label{eq:sigmavelavg}
\end{align}
In principle, we should include a Sommerfeld enhancement correction. 
The correction starts becoming important only when $\pi \alpha_{\rm EM} \epsilon^2 (m_\chi\,c^2/4\,\kb\, T_\chi)^{1/2} \simeq 1$, which is never achieved for parameters of interest.

Pair annihilation is inefficient in depleting the DM at low redshifts. 
However, each annihilation releases an amount of energy that is large compared to the average energy per particle during cosmic dawn, so in principle we should include an additional heating. 
If we have $m_\chi \gg {\rm max}(\me , \kb\,T_\chi/c^2)$, then the dominant source of energy is the rest mass of the DM particles, which converts mostly into the kinetic energy of the electrons and then thermalize locally.

The heating efficiency due to DM annihilation is 
\begin{align}
  \Edma & = \frac{\fdm^2 n_\chi^2 \langle \sigma_{\rm a} v_{\rm rel} \rangle m_\chi c^2}{3 \nh H \kb \Tk} \label{eq:qbdotann} \\
  & \approx 5.6 \times 10^{-3} \left( \frac{\fdm}{10^{-2}} \right)^2 \left( \frac{\epsilon}{10^{-5}} \right)^2 \left( \frac{m_\chi}{10 \, {\rm MeV}} \right)^{-3} \times \notag \\
  &~~~ \left( \frac{\Tk}{7 \, {\rm K}} \right)^{-1} \left( \frac{1+z}{18} \right)^{3/2} \sqrt{1 - \frac{\me^2}{m_\chi^2}} \left( 1 + \frac{\me^2}{2 m_\chi^2} \right).
\end{align}
When computing the evolution of $\Tk$, we just add the terms in Eqs.~\eqref{eq:qbdot} and \eqref{eq:qbdotann} inside the brackets in Eq.~\eqref{eq:tevol}. 
Apart from a factor of two in Eq.~\eqref{eq:qbdotann} and an additional square root in Eq.~\eqref{eq:sigmavelavg}, our final equations are identical to those in Ref.~\cite{2018arXiv180309739L}. 

What are the appropriate boundary conditions for integrating Eqs.~\eqref{eq:dvchibdz} and \eqref{eq:dtchidz}?
MKA15 start their integrations at the redshift of kinetic decoupling, $z = 1010$, where they take $T_\chi = 0$ and sample $\Vchib$ from the \lcdm \, distribution.
This procedure is valid at the low values of the cross sections they considered ($\sigma_{0,t} \sim 10^{-41} \, {\rm cm}^2$), while the typical values used to explain the EDGES results are much larger. 
For such cross sections, the temperature of the DM can satisfy $T_\chi \approx \Tk$ at higher redshifts, and the relative velocity $\Vchib$ can be quickly damped, so that the fraction of charged DM effectively behaves like an extra baryonic component (this is the strong-coupling regime of Ref.~\cite{2018arXiv180309739L}).
For the parameters considered in the main text, we start integrations with $T_\chi = 0$ at a sufficiently high redshift before recombination, at which point $T_\chi$ is quickly set to $\approx \Tk$.
Note also that the coupled differential equations, \eqref{eq:dvchibdz} and \eqref{eq:dtchidz}, can be stiff during the epoch of decoupling of the DM, and thus must be integrated with care.


\bibliographystyle{apsrev4-1-etal}
\bibliography{references}

\end{document}